\documentclass[twocolumn,showpacs,preprintnumbers,aps]{revtex4}
\newcommand{\BE}{\begin{equation}}
\newcommand{\EE}{\end{equation}}
\newcommand{\BA}{\begin{eqnarray}}
\newcommand{\EA}{\end{eqnarray}}
\usepackage{graphicx}
\usepackage{dcolumn}
\usepackage{bm}
\input epsf.tex

\begin{document}

\title{Diffusive and localization behavior of electromagnetic waves in a two-dimensional random medium}

\author{Ken Kang-Hsin Wang} \author{Zhen
Ye}\email{zhen@phy.ncu.edu.tw} \affiliation{Wave Phenomena
Laboratory and Center of Complex Systems, Department of Physics,
National Central University, Chungli, Taiwan 32054, Republic of
China}

\date{\today}

\begin{abstract}

In this paper, we discuss the transport phenomena of
electromagnetic waves in a two-dimensional random system which is
composed of arrays of electrical dipoles, following the model
presented  earlier by Erdogan, et al. (J. Opt. Soc. Am. B {\bf
10}, 391 (1993)). A set of self-consistent equations is presented,
accounting for the multiple scattering in the system, and is then
solved numerically. A strong localization regime is discovered in
the frequency domain. The transport properties within, near the
edge of and nearly outside the localization regime are
investigated for different parameters such as filling factor and
system size. The results show that within the localization regime,
waves are trapped near the transmitting source. Meanwhile, the
diffusive waves follow an intuitive but expected picture. That is,
they increase with travelling path as more and more random
scattering incurs, followed by a saturation, then start to decay
exponentially when the travelling path is large enough, signifying
the localization effect. For the cases that the frequencies are
near the boundary of or outside the localization regime, the
results of diffusive waves are compared with the diffusion
approximation, showing less encouraging agreement as in other
systems (Asatryan, et al., Phys. Rev. E {\bf 67}, 036605 (2003).)

\end{abstract}

 \maketitle

\section{Introduction}

Waves are a ubiquitous phenomenon that is relevant to our
everyday's life. Our knowledge about the Nature is mainly obtained
through either acoustic or electromagnetic waves. When waves
propagate through a random medium, some peculiar properties would
emerge and some of them have stood as a long outstanding problem
for physicists. The most intriguing and still unsolved puzzle
perhaps is the so called Anderson localization. The concept of
Anderson localization was initially proposed nearly half a century
ago for the possible phenomenon of disorder-induced
metal-insulator transition in electronic systems \cite{Anderson}.
To make it simple, Anderson localization refers to situations that
when released in a random medium which could be, for instance, a
free space with random potentials, electrons may go nowhere but
stay close to the initial place. The envelop of the electronic
wave function subsequently reveals as an exponential decay along
any direction from the emitting point \cite{Lee}; the length
measuring the e-fold decay is called the localization length. The
mechanism behind this property has been attributed purely to
sufficient multiple scattering of electrons by the random
potentials, a feature of the wave nature of electrons.

Since its inception, the localization concept has opened a wide
door for scientists from various backgrounds, and stimulated a
vast body of research. The fact that electronic localization is
due to the wave nature of electrons is particularly important, as
the concept may also be applied to classical wave systems by
analogy \cite{Maynard}. The concept of localization has far
reached many fields such as seismology\cite{Seis},
oceanology\cite{ocean}, and random lasers\cite{Laser}, to name
just a few. The great efforts have been summarized in a number of
reviews (e. g.
\cite{Lee,Thouless,John,Sheng,Lagen,Rossum,Van,loc}).

Over the past two decades, localization of electromagnetic and
acoustic waves has been and continued to be a particularly
attracting problem, leading to a great amount of publications
(e.~g.
\cite{John,Sheng,Lagen,HW,Kirk1,He,Condat,Sor,Genack,McCall,McCall2,Marian,Sigalas,AAA,Ye1,Wiersma,weak,AAC}).
The theory of classical waves has been detailed in the
textbook\cite{Sheng}. In spite of the tremendous efforts, however,
there are still some open questions for wave localization in two
dimensional (2D)  random systems. For example, one of the
questions concern with the transport behavior of waves within and
outside localization regimes. It has been discussed much in the
literature that within the localization regime the wave intensity
follows a diffusive behavior for transport path smaller than
localization length, whereas it follows an exponential decay when
the path is greater than the localization length
\cite{Sheng,Zhang}. In addition, it has been conjectured that all
waves are bound to be localized in 2D for any given amount of
disorders or randomness\cite{gang4}, a conclusion significantly
influencing nearly all later investigations. If this general
hypothesis is true, one would expect that the wave intensity would
have to follow the diffusion to localization transition as the
transport distance increases although such a transition has not
been solidly confirmed yet. Recent studies, however, tended to
suggest that this conjecture may be generally invalid
\cite{Ye2,diffusion}. One one hand, in \cite{Ye2}, it has been
suggested that acoustic waves could not be localized in certain
frequency ranges. One the other, the research reported in
\cite{diffusion} shows different frequency regimes of
electromagnetic wave transport. In certain ranges, the wave
transport behavior can be entirely described by the usual
Boltzmann diffusion approximate theory, while in other ranges the
behavior does not reveal a diffusive feature. The authors called
the latter phenomenon as anomalous diffusion, and inferred as the
incipience of Anderson localization. These works seem to indicate
that as the frequency varies, there will be a transition between
localization and non-localization even in two dimensions. Whether
these observations also hold in a general perspective remains to
be known.

The difficulties in the study of possible non-localization or
localization of EM waves mainly lie in a number of problems.
First, wave localization only appears for strongly scattering
media, and such a medium is often hard to find. Second,
localization effects are often entangled with other effects such
as dissipation, wave deflection, or boundary effects\cite{loc},
making data interpretation often ambiguous. Third, many two
dimensional systems are exactly solvable by numerical computation.
But the numerical simulation is very time-consuming and is
obviously limited by computing facilities with regard to some
unavoidable problems such as finite size. How to find a suitable
model that could ease these concerns poses a challenge problem in
its own right.

In a recent communication, a simple but seemingly realistic model
system has been proposed to study EM localization in 2D random
media\cite{YLS}. This model originated from the previous study of
the radiative effects of the electric dipoles embedded in
structured cavities\cite{Erdogan}. It was shown that EM
localization is possible in such a disordered system. When
localization occurs, a coherent behavior appears and is revealed
as a unique property differentiating localization from either the
residual absorption or the attenuation effects.

With the present paper, we wish to explore further the transport
properties of the system outlined in \cite{YLS}. The advantage of
this system not only relies on its simplicity, but on its less
time consumption. It has been shown that there is a single strong
localization region in the system. We will investigate the
transport behavior of wave intensity within, near or outside the
strong localization regime. We study how the transport behavior
depends on scatterer's filling parameter, frequency, and sample
size and so on. When the localization effect is less obviously, we
compare the numerically evaluated wave intensity with the result
from the the diffusion theory.

The paper is structure as follows. The description of the system
and the relevant theoretical modelling are presented in next
section, followed by the numerical simulation and detailed
discussion of results. A summary concludes the paper in the last
section.

\section{The system and theoretical formulation}

Here we present the system and the theoretical formulation.
Although these have already been presented before \cite{Wang}, for
the sake of the convenience on the reader's part and easy
discussion on our part, we purposely repeat the essence here.

\subsection{The system}

Following Erdogan et al.\cite{Erdogan}, we consider 2D dipoles as
an ensemble of harmonically bound charge elements. In this way,
each 2D dipole is regarded as a single dipole line, characterized
by the charge and dipole moment per unit length. Assume that $N$
parallel dipole lines, aligned along the $z$-axis, are embedded in
a uniform dielectric medium and {\it randomly} located at
$\vec{r}_i (i=1,2,\dots,N)$. The cross section of the dipoles are
assumed to be in the $x-y$ plane. The averaged distance between
dipoles is $d$. A stimulating dipole line source is located at
$\vec{r}_s$, transmitting a continuous wave of angular frequency
$\omega$. By the geometrical symmetry of the system, we only need
to consider the $z$ component of the electrical waves.

\subsection{The formulation}

Although much of the following materials can be referred to in
\cite{YLS}, we repeat the important parts here for the sake of
convenience and completeness.

Upon stimulation, each dipole will radiate EM waves. The radiated
waves will then repeatedly interact with the dipoles, forming a
process of multiple scattering. The equation of motion for the
$i$-th dipole is \BA \frac{d^2}{dt^2}p_i + \omega_{0,i}^2p_i &=&
\frac{q_i^2}{m_i}E_{z}(\vec{r}_i) - b_{0,i}\frac{d}{dt}
p_i,\nonumber\\
& & \mbox{for} \ i = 1, 2,\dots, N. \label{eq:1}\EA where
$\omega_{0,i}$ is the resonance (natural) frequency, $p_i$, $q_i$
and $m_i$ the dipole moment, charge and effective mass per unit
length of the $i$-th dipole respectively. $E_{z}({\vec{r}_i})$ is
the total electrical field acting on dipole $p_i$, which includes
the radiated field from other dipoles and also the directly field
from the source. The factor $b_{0,i}$ denotes the damping due to
energy loss and radiation, and can be determined by energy
conservation. Without energy loss (to such as heat), $b_{0,i}$ can
be determined from the balance between the radiative and
vibrational energies, and is given as\cite{Erdogan} \BE b_{0,i} =
\frac{q_i^2\omega_{0,i}}{4\epsilon m_i c^2}, \label{eq:2}\EE with
$\epsilon$ being the constant permittivity and $c$ the speed of
light in the medium separately.

Equation (\ref{eq:1}) is virtually the same as Eq.~(1) in
\cite{Erdogan}. The only difference is that in \cite{Erdogan},
$E_z$ is the reflected field at the dipole due to the presence of
reflecting surrounding structures, while in the present case the
field is from the stimulating source and the radiation from all
other dipoles.

The transmitted electrical field from the continuous line source
is determined by the Maxwell equations\cite{Erdogan}\BE \left
(\nabla^2 - \frac{\partial^2}{c^2\partial t^2}\right
)G_0(\vec{r}-\vec{r}_s) =
-4\mu_0\omega^2p_0\pi\delta^{(2)}(\vec{r}-\vec{r}_s) e^{-i\omega
t}, \label{eq:3}\EE where $\omega$ is the radiation frequency, and
$p_0$ is the source strength and is set to be unit. The solution
of Eq.~(\ref{eq:3}) is clearly \BE G_0(\vec{r}-\vec{r_s}) =
(\mu_0\omega^2p_0) i\pi H_0^{(1)}(k|\vec{r}-\vec{r}_s|)
e^{-i\omega t}, \label{eq:4}\EE with $k=\omega/c$, and $H_0^{(1)}$
being the zero-th order Hankel function of the first kind.

Similarly, the radiated field from the $i$-th dipole is given by
\BE \left (\nabla^2 - \frac{\partial^2}{c^2\partial t^2}\right
)G_i(\vec{r}-\vec{r}_i) =
\mu_0\frac{d^2}{dt^2}p_i\delta^{(2)}(\vec{r}-\vec{r}_i).
\label{eq:5}\EE The field arriving at the $i$-th dipole is
composed of the direct field from the source and the radiation
from all other dipoles, and thus is given as \BE E_z(\vec{r}_i) =
G_0(\vec{r}_i - \vec{r}_s) + \sum_{j=1, j\neq i}^N G_j(\vec{r}_i -
\vec{r}_j). \label{eq:6}\EE

Substituting Eqs.~(\ref{eq:4}), (\ref{eq:5}), and (\ref{eq:6})
into Eq.~(\ref{eq:1}), and writing $p_i = p_ie^{-i\omega t}$, we
arrive at \BA (-\omega^2 + \omega^2_{0,i} - i\omega b_{0,i})p_i =&
&\nonumber\\ \frac{q_i^2}{m_i}\left[G_0(\vec{r}_i - \vec{r}_s) +
\sum_{j=1, j\neq i}^N \frac{\mu_0 \omega^2}{4}
iH_0^{(1)}(k|\vec{r}_i-\vec{r}_j|)p_j\right].& &  \label{eq:7}\EA
This set of linear equations can be solved numerically for $p_i$.
After $p_i$ are obtained, the total field at any space point can
be readily calculated from \BE E_z(\vec{r}) = G_0(\vec{r} -
\vec{r}_s) + \sum_{j=1}^N G_j(\vec{r} - \vec{r}_j).\label{eq:8}\EE

In the standard approach to wave localization, waves are said to
be localized when the square modulus of the field
$|E(\vec{r})|^2$, representing the wave energy, is spatially
localized after the trivial cylindrically spreading effect is
eliminated. Obviously, this is equivalent to say that the further
away is the dipole from the source, the smaller its oscillation
amplitude, expected to follow an exponentially decreasing pattern.

An alternative two dimensional dipole model was devised previously
\cite{Marian}. The authors derived a set of linear algebraic
equations, which is similar in form to the above Eq.~(\ref{eq:6}).
However, the fundamental difference between the two models is with
respect to the relation between the incident and scattered waves.
In \cite{Marian}, the interaction between dipoles and the external
field is derived by the energy conservation, while in the present
case the coupling is determined without ambiguity by the Newton's
second law. The former leads to an undetermined phase factor.
According to, e.~g. Refs.~\cite{Erdogan,ccw}, the energy
conservation can only give the radiation factor in
Eq.~(\ref{eq:2}).

There are several ways to introduce randomness to
Eq.~(\ref{eq:7}). For example, the disorder may be introduced by
randomly varying such properties of individual dipoles as the
charge, the mass or the two combined. This is the most common way
that the disorder is introduced into the tight-binding model for
electronic waves\cite{Zallen}. In the present study, the disorder
is brought in by the random distribution of the dipoles.

For simplicity yet without losing generality, assume that all the
dipoles are identical and they are randomly distributed within a
square area. The source is located at the center (set to be the
origin) of this area. For convenience, we make Eq.~(\ref{eq:7})
non-dimensional by scaling the frequency by the resonance
frequency of a single dipole $\omega_0$. This will lead to two
independent non-dimensional parameters $b = \frac{q^2\mu_0}{4m}$
and $b_0^\prime =
\frac{\omega}{\omega_0}\left(\frac{b_0}{\omega_0}\right)$. Both
parameters may be adjusted in the experiment. For example, the
factor $b_0$ can be modified by coating layered structures around
the dipoles\cite{Erdogan}. Then Eq.~(\ref{eq:7}) becomes simply
\BA (-f^2+1 -ib_0^\prime)p_i =
& & \nonumber\\
ibf^2\left[p_0H_0^{(1)}(k|\vec{r}_i-\vec{r}_s|) + \sum_{j=1, j\neq
i}^N p_jH_0^{(1)}(k|\vec{r}_i-\vec{r}_j|)\right] && \label{eq:10})
\EA with $f=\frac{\omega}{\omega_0}$. Eq.~(\ref{eq:10}) is
self-consistent and can be solved numerically for $p_i$ and then
the total field is obtained through Eqs.~(\ref{eq:3}),
(\ref{eq:5}) and (\ref{eq:8}).

\subsection{Expected transport behavior}

\subsubsection{In general}

Following \cite{3D}, a general consideration of the spatial
behavior of wave transport is possible. Consider a wave
transmitted in a random medium. The transport equation for the
total energy intensity $I$, i.~e. $<|E|^2>$, may be intuitively
written as \BE \frac{dI}{dx} = -\alpha I, \label{eq:1a} \EE where
$\alpha$ represents decay along the path traversed. After
penetrating into the random medium, the wave will be scattered by
random inhomogeneities. As a result, the wave coherence starts to
decrease, yielding the way to incoherence. Extinction of the
coherent intensity $I_C$, i.~e. $|<E>|^2$, is described by \BE
\frac{dI_C}{dx} = -\gamma I_C, \label{eq:2a} \EE with the
attenuation constant $\gamma$. Eqs.~(\ref{eq:1a}) and
(\ref{eq:2a}) lead to the exponential solutions \BE I(x) =
I(0)e^{-\alpha x}, \ \ \ \mbox{and} \ \ \ I_C(x) = I(0)e^{-\gamma
x}. \EE In deriving these equations, the boundary condition was
used; it states that $I(0) = I_C(0)$ as no scattering has been
incurred yet at the interface. According to energy conservation,
the incoherent intensity $I_D$ (diffusive) is thus \BE I_D(x) =
I(x) - I_C(x). \EE

When there is no absorption, the decay constant $\alpha$ is
expected to vanish and the total intensity will then be constant
along the propagation path. Then, the coherent energy gradually
decreases due to random scattering and transforms to the diffusive
energy, while the sum of the two forms of energy remains a
constant. This scenario, however, changes when localization
occurs. Even without absorption, the total intensity can be
localized near the interface due to multiple scattering. When this
happens, $\alpha$ does not vanish. The transport of the total
intensity may be still described by Eq.~(\ref{eq:1a}), and the
inverse of $\alpha$ would then refer to the localization length.

\subsubsection{Diffusion theory}

As discussed above, as waves propagate, the coherent energy
gradually decreases and the incoherent (the so called diffusive
waves) progressively grows. In the absence of localization, the
diffusive intensity (or energy) will eventually surpasses the
coherent energy at a certain point, then becomes the dominating
component in the wave transmission. This intuitive physical
picture has been nicely phrased in the very excellent textbook by
Ishimaru \cite{Ishimaru}.

The diffusive waves are expected to follow the diffusion equation
which can be described by \BE \left(D\nabla^2 -
\frac{\partial}{\partial t}\right) I_D =
-S\delta(\vec{r}-\vec{r}_s),\EE where $D$ is the diffusive
constant and $S$ denotes a strength factor which is related to the
total emitted energy ($I_0$) and transport mean free path ($l_t$).
In the steady situation, this equation is simplified as
\cite{Ishimaru} \BE \nabla^2 I_D =
-\frac{2I_0}{l_t}\delta(\vec{r}-\vec{r}_s).\EE The boundary
condition for the diffusive waves is \cite{Ishimaru} \BE \left.
I_D + \frac{2l_t}{3}\frac{\partial I}{\partial n}\right|_{n} =
0,\EE with $\vec{n}$ being the outward normal at the boundary.

As discussed in \cite{Chen2002}, the traditional wave of measuring
for localization effects, that is putting the source at one side
of a sample and measuring the transmission across the sample on
the other side, may be influenced rather significantly by other
non-localization effects such as reflection, in the present paper
we use an alternative setup, in line with the previous studies
\cite{diffusion,Chen2002}: Put the source in the middle of a
scattering sample which takes a circular shape. In this case, a
solution to the diffusive intensity is given as  \cite{diffusion}
\BE I_D = - \frac{I_0}{\pi l_t}\ln(r/R) + I_b, \label{eq:d}\EE
where $R$ is the samole radius, i.~e. the size of the cluster of
the scatterers, and $I_b$ is the intensity at the boundary. In
practice, $l_t$ can be regarded as a free parameter to fit the
numerical simulation. When the diffusive energy is dominant, the
total energy would be expected to also follow the behavior
characterized by Eq.~(\ref{eq:d}).

\section{The results and discussion}

\begin{figure}[hbt]
\vspace{10pt} \epsfxsize=2.5in\epsffile{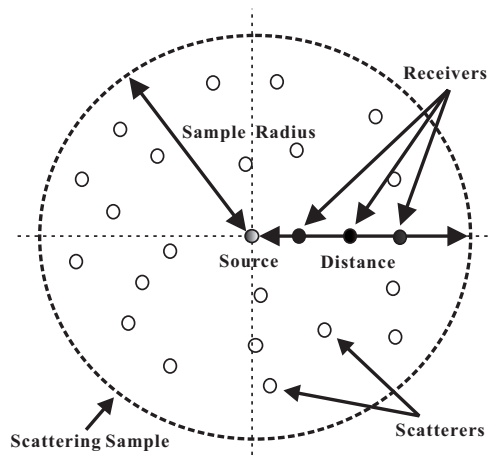}
\caption{Conceptual layout of the system and the simulation.}
\label{fig1}
\end{figure}

A conceptual set up of the system for discussion is pictured by
Fig.~\ref{fig1}. Here it is shown that a source is located at the
center of a sample circled by the broken line. The small circles
inside refer to the dipoles as the scatterers. The receiver is
placed at various spatial positions along the radial direction to
record the transmitted waves.

\subsection{Parameters in the simulation}

Unless otherwise noted, the following parameters are used in the
numerical simulation: the non-dimensional damping rate,
$b_0/\omega_0 = 0.001$ and the interaction coupling, $b = 0.001$.
Without notification, the filling factor ($\beta$) is taken as
6.25; the filling factor is defined as the number of dipoles per
unit area. The number of random configurations for averaging is
taken in such a way that the convergency is assured. In the
calculation, we scale all lengths by a length $l$ such that
$k_0l=1$, and frequency by $\omega_0$. In this way, the frequency
always enters as $k/k_0$. We find that all the results shown below
are only dependent on parameters $b$, $b_0/\omega_0$, and the
ratio $\omega/\omega_0$ or equivalently $k/k_0$. Such a simple
scaling property may facilitate designing experiments. In the
numerical computation, we take $c=1$ for convenience. The total
normalized wave at a spatial point is scaled as $T(\vec{r}) \equiv
E(\vec{r})/E_0(\vec{r})$, with $E_0$ being the direct wave from
the source, so that the trivial geometric spreading effect is
naturally removed. When compared with diffusion theory, the
diffusive energy would have to be multiplied by the spreading
factor, and then is $I_Dr$.

\subsection{Numerical results and discussion}

\begin{figure}[hbt]
\vspace{10pt} \epsfxsize=2.5in\epsffile{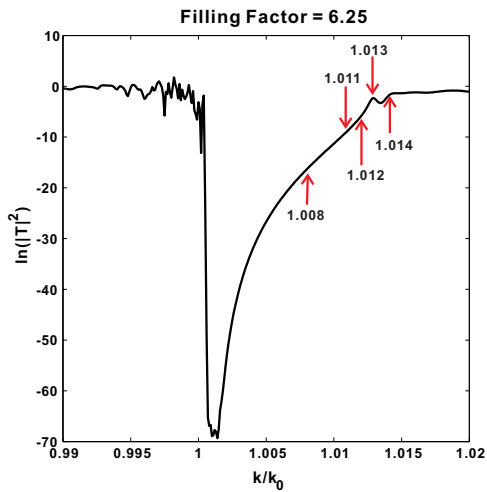} \caption{The
transmission versus the frequency.} \label{fig2}
\end{figure}

First we plot the wave transmission versus frequency to locate the
strong localization region. The result is shown by
Fig.~\ref{fig2}. Here totally 1964 dipoles are considered,
amounting to a sample radius of about 10. The receiver is located
outside at about 1.2 radius of the sample. We see that there is a
very strong inhibition region ranging roughly from 1.001 up to
1.012. Outside this regime, the inhibition is not obvious. We have
tried our current computing facility to its extremes with the
largest possible number of dipoles, but we still failed to see
obvious inhibition outside the regime. The width of the strong
inhibition window mainly depends on the the filling factor. We
choose five frequencies, indicated on Fig.~\ref{fig2}, to study
the localization or diffusion behaviors of the wave intensity.

\begin{figure}[hbt]
\vspace{10pt} \epsfxsize=2.5in\epsffile{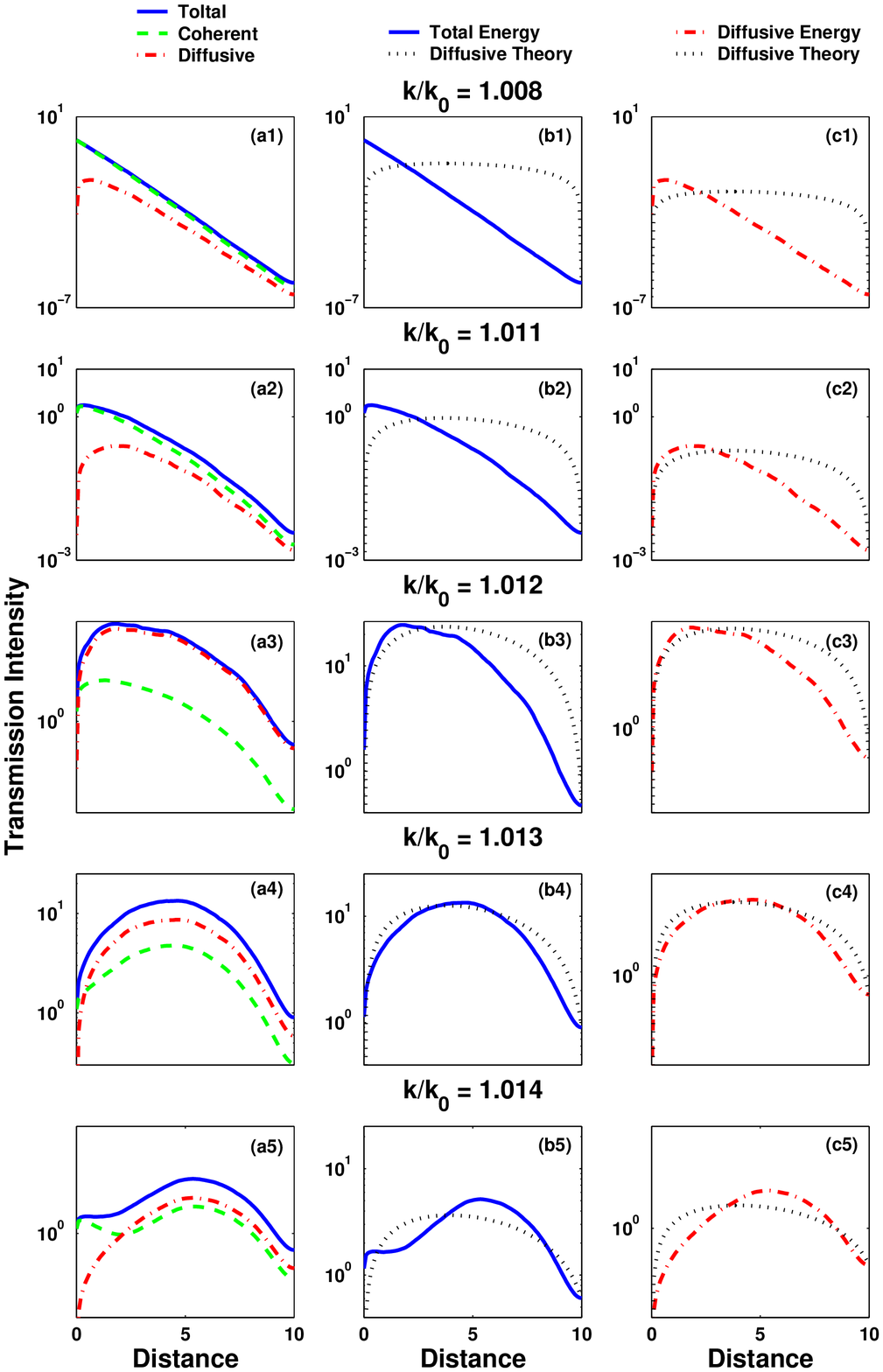}
\caption{Transport behavior: (a) the total, coherent and diffusive
intensity as a function of travelling distance; (b) the comparison
of the numerical results of the total energy with that from the
diffusion theory; (c) the comparison of the numerical results of
the diffusive energy with that from the diffusion theory. We
considered 1964 dipoles to form a sample whose radius is roughly
10.} \label{fig3}
\end{figure}

In Fig.~\ref{fig3}, we plot the transmission behavior of various
intensities as a function of wave travelling distance. Enough
average over the random configurations of the scatterers is taken
to ensure the stability of the results. The maximum number of the
average is 1200. The observation can be summarized as follows. (1)
It is clear from the left panel that within the strong inhibition
window, the transmission follows perfectly the expected
localization behavior; the total and coherent intensity decays
exponentially with slight different slopes respectively, while the
diffusive intensity increases initially, then saturates, followed
by a decay afterwards. These are shown by (a1). (2) As the
frequency moves towards the edge of the strong inhibition window,
the behavior starts to change more and more severely. As the
frequency increases gradually, the total energy starts to behave
more like the diffusive waves, although the diffusive waves are
not yet dominant (such as in (a2)). But overall speaking, the
transport behavior of the total energy behaves more like the
diffusive waves when the latter are prominent, as expected and
shown by (a3) and (a4). (4) At frequencies for which the diffusive
intensity is dominant, the behavior of the diffusive energy is
qualitatively in agreement with the prediction of the diffusion
theory, such as in the cases described by (c3), but in certain
cases the agreement is good both qualitatively and quantitatively
as seen in (c4). Overall speaking, however, the agreement is less
encouraging compared with the simulation for another 2D system
illustrated in \cite{diffusion}. And the rapid fluctuation of the
diffusive energy shown in \cite{diffusion} is absent here. Another
note is worth making here. In \cite{diffusion}, the authors also
compared the results within a strong localization region. The
exponential decay revealed in their figures, for example Fig.~9,
is only possible when the cylindrical geometrical spreading factor
$1/r$ is taken out. It seems to us that this factor has not taken
into account in their discussion. (5) Just by looking, we are
tempted to conclude that there is indeed a fundamental difference
in the transport behaviors between within and outside the strong
inhibition region. This conclusion accords qualitatively with
\cite{diffusion}. (6) We also observe that outside the strong
inhibition regime, the coherent intensity dose not necessarily
behave as an exponential decay, exemplified by (a3) and (a4). This
is in contrast to usual expectation. A plausible cause is the
finite size effects. In this situation, the mean free path is hard
to estimate, unlike in other systems \cite{Zhang,Chen2002}.

\begin{figure}[hbt]
\vspace{10pt} \epsfxsize=2.5in\epsffile{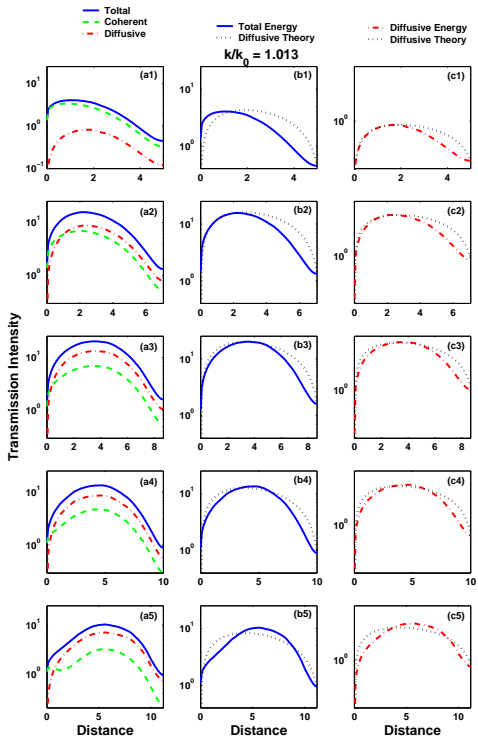}
\caption{Dependence on the sample size of the transport behavior.
The following parameters are taken. (1). k/k0 = 1.013; (2).
Filling factor = 6.25; (3). The number of average: 400; (4).
Fig(a1 to c1) : dipole number: 491, Sample radius: 5; Fig(a2 to
c2): dipole number: 982, Sample radius: 7; Fig(a3 to c3) : dipole
number: 1473, Sample radius: 9; Fig(a4 to c4) : dipole number:
1964, Sample radius: 10; Fig(a5 to c5) : dipole number: 2455,
Sample radius: 11.} \label{fig4}
\end{figure}

\begin{figure}[hbt]
\vspace{10pt} \epsfxsize=2.5in\epsffile{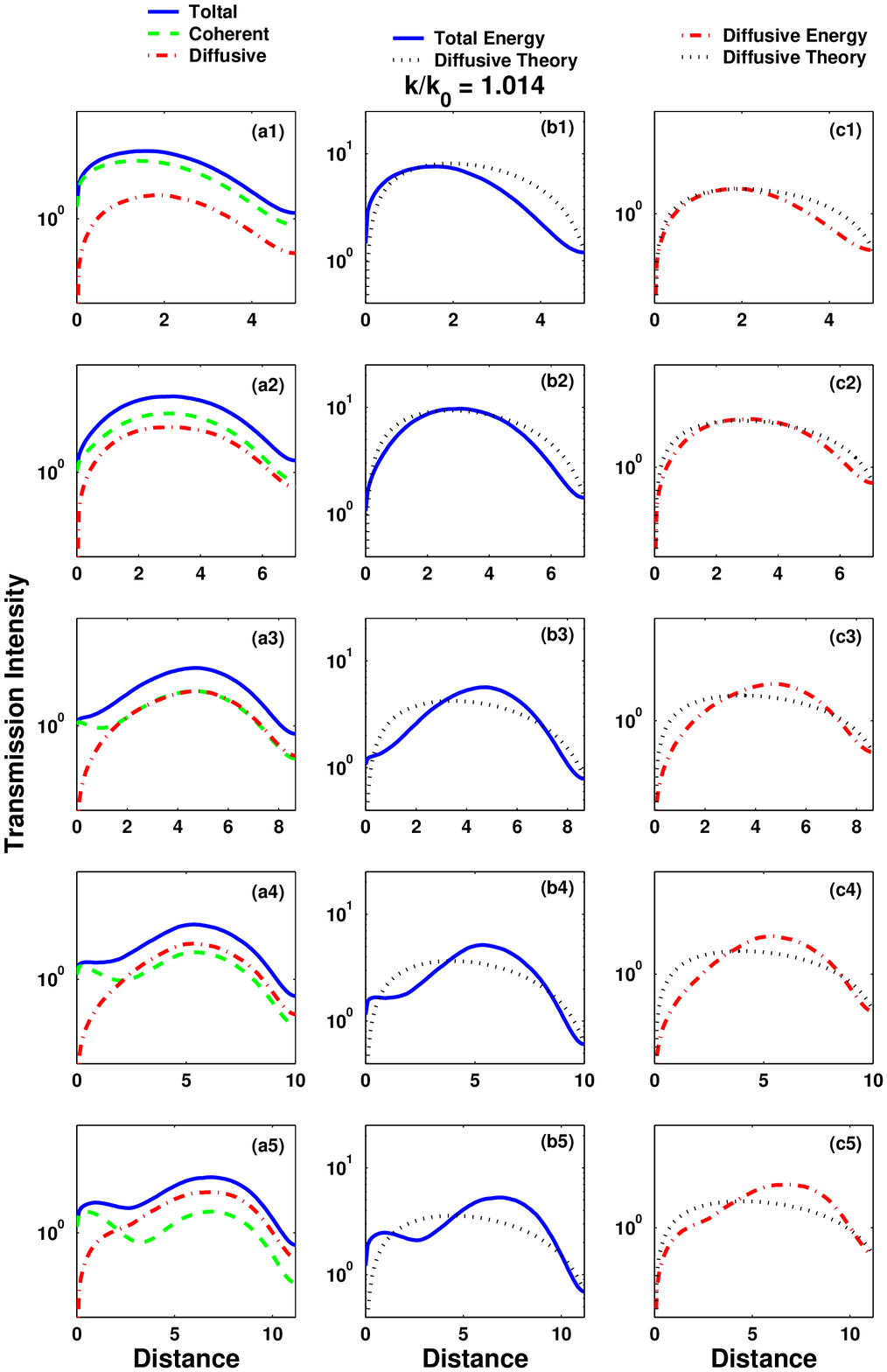}
\caption{Dependence on the sample size of the transport behavior.
The following parameters are taken. (1). k/k0 = 1.014; (2).
Filling factor = 6.25; (3). The number of average: 400; (4).
Fig(a1 to c1) : dipole number: 491, Sample radius: 5; Fig(a2 to
c2): dipole number: 982, Sample radius: 7; Fig(a3 to c3) : dipole
number: 1473, Sample radius: 9; Fig(a4 to c4) : dipole number:
1964, Sample radius: 10; Fig(a5 to c5) : dipole number: 2455,
Sample radius: 11.} \label{fig5}
\end{figure}

From Fig.~\ref{fig3}, we see that at around $k/k_0 = 1.013$, just
outside the upper edge of the strong localization boundary, the
diffusive waves are dominant and the transport behavior tends to
follow the prediction from the diffusion theory. We would like now
to explore the robustness of such an agreement. The results are
shown in Fig.~\ref{fig4}. Here it is shown that as the size of the
sample is increased, the diffusive intensity become more and more
prominent and meanwhile the agreement between the diffusion and
exact numerical results is improved, as shown by the diagrams from
the top downwards. When we consider another frequency $k/k_0 =
1.014$, the agreement seems not improving as the sample size is
enlarged, although the two frequencies are close. The comparison
is shown in Fig.~\ref{fig5}. The best agreement is when the sample
size is 7, referring to (b2) and (c2). From these results, one may
say that the agreement between the diffusion theory and the exact
numerical results may not be robust, at least in the present
system. We have to note that the results from the diffusion theory
are approximate. The deviation from the diffusion theory is
therefore not out of expectation.

\begin{figure}[hbt]
\vspace{10pt} \epsfxsize=2in\epsffile{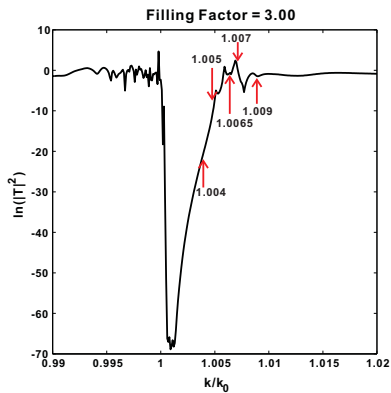} \caption{The
transmission versus the frequency. The parameters are as follows.
(1) 2150 dipoles in a circular sample, amounting to a sample
radius of 15; (2) the filling factor = 3.00; (3) the receiver is
positioned at 1.2 radius of the sample.} \label{fig6}
\end{figure}

We also considered the case for another filling factor at 3.00.
The general features are agreeable with the above case with 6.26.
For brevity, we just show partial results. Fig.~\ref{fig6} shows
the transmission versus the frequency. The parameters are shown in
the figure caption. Compared to the case with filling factor of
6.25, we see that decreasing filling factor only tends to reduce
the strong inhibition regime while all other features remain
unchanged.

\begin{figure}[hbt]
\vspace{10pt} \epsfxsize=2.5in\epsffile{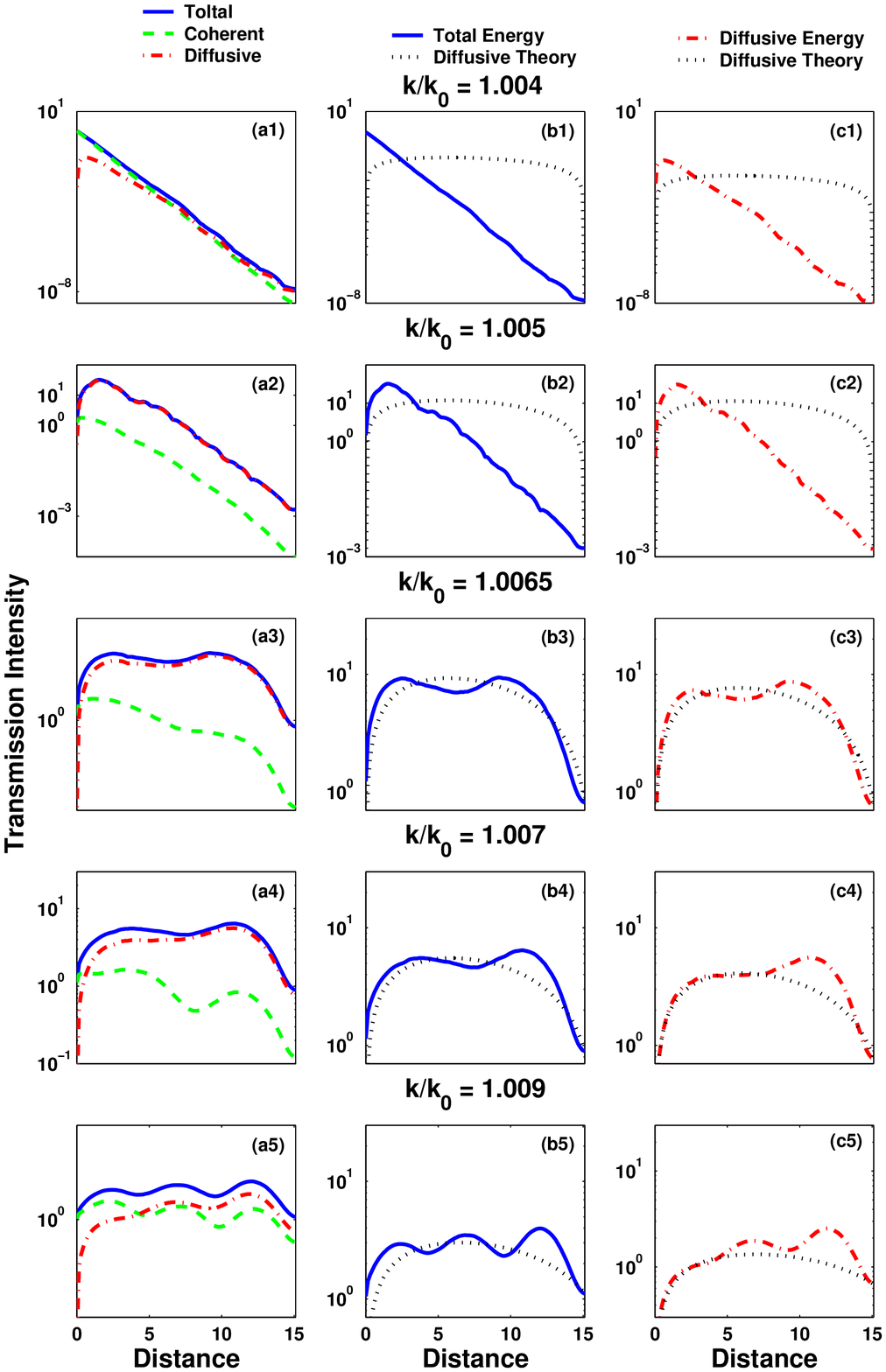}
\caption{Transport behavior: (a) the total, coherent and diffusive
intensity as a function of travelling distance; (b) the comparison
of the numerical results of the total energy with that from the
diffusion theory; (c) the comparison of the numerical results of
the diffusive energy with that from the diffusion theory. We
considered 2150 dipoles to form a sample whose radius is roughly
15.} \label{fig7}
\end{figure}

Again we choose five frequencies to inspect the transport
behavior, and show the results in Fig.~\ref{fig7}. The general
conclusions drawn from Fig.~\ref{fig3} remain. However, there are
differences. For the frequency 1.004 which is within the strong
localization regime, the transport behavior fully complies with
the expectation. The diffusive intensity increases, saturates, and
then decreases to take over as the dominating part in the total
transmission. We re-plot fig.~\ref{fig7} (a1) in Fig.~8 to show
this behavior. The cross over point when the diffusive part
becomes dominating is represented by the vertical line in the
figure.

\begin{figure}[hbt]
\vspace{10pt} \epsfxsize=1.5in\epsffile{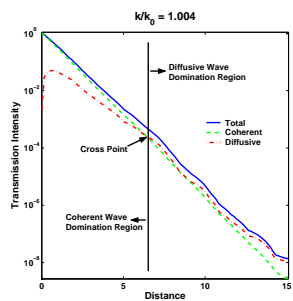}
\caption{Re-plotted results in Fig.~\ref{fig7}(a1).} \label{fig8}
\end{figure}

\begin{figure}[hbt]
\vspace{10pt} \epsfxsize=2.in\epsffile{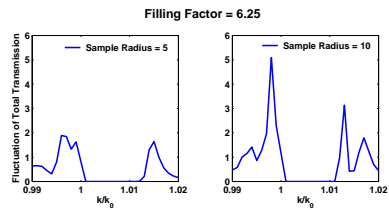} \caption{The
fluctuation of the normalized transmission intensity versus
frequency, i.~e. $\delta T^2 = <T^2> - <T>^2$.} \label{fig9}
\end{figure}

Lastly, we have computed the fluctuation of the normalized
transmission intensity versus frequency. The results are shown in
Fig.~\ref{fig9}. The parameters are as follows: filling factor
(Number of dipole/area) = 6.25;  two sample sizes are 5 and 10
respectively; the receiver is at 1.2 times the radius; the number
of average is 100. The results indicate that the fluctuation is
nearly zero with the localization regions. Outside these regimes,
the fluctuation is strong. The results also show that the
localization regime is the same for different sample sizes. These
features hint that there may have a localization transition in 2D
random media, in accordance with the previous results
\cite{Ye2,diffusion}.

\section{Summary}

In this article, the transport properties of electromagnetic waves
in a two-dimensional random system. Some general properties of the
transport phenomena are elaborated. For certain ranges of
frequencies, strongly localized electromagnetic waves have been
observed in such a simple but realistic disordered system. The
spatial behavior of the total, coherent and diffusive waves is
explored within, near the edge of and outside the localization
regime, and are investigated for different parameters such as
filling factor and system size. The results show that within the
localization regime, waves are trapped near the transmitting
source. Meanwhile, the diffusive waves follow an intuitive but
expected picture. For the cases that the frequencies are near the
boundary of or outside the localization regime, the results of
diffusive waves are compared with the diffusion approximation,
showing less encouraging agreement as in other systems
\cite{diffusion}. Furthermore, the results tend to suggest that
the study of the fluctuation behavior in the transmission may also
help identify the localization regime.

\section*{Acknowledgments}

This work is supported by the National Science Council of Republic
of China (Grant No. 91-2112-M008-024). Supports from Department of
Physics, College of Science, and the National Central University
are also greatly appreciated. Without these helps, the simulation
carried out here would be impossible.

\end{document}